\documentclass[fleqn,usenatbib]{mnras}

\usepackage{newtxtext,newtxmath}

\usepackage[T1]{fontenc}
\usepackage{ae,aecompl}

\usepackage{graphicx}	
\usepackage{amsmath}	
\usepackage{amssymb}	

\title[Visualization for Virtual Observatories]{Interactive 3D Visualization for Theoretical Virtual Observatories}

\author[T. Dykes et al.]{
Tim Dykes, $^{1}$\thanks{E-mail: tim.dykes@port.ac.uk} 
A. Hassan,$^{2}$
C. Gheller,$^{3}$
D. Croton,$^{2}$
M. Krokos,$^{1}$
\\
$^{1}$School of Creative Technology, University of Portsmouth, Eldon Building, Winston Churchill Ave, Portsmouth PO1 2UP, UK\\
$^{2}$Centre for Astrophysics and Supercomputing, Swinburne University of Technology, Hawthorn, Victoria, 3122, Australia\\
$^{3}$CSCS-ETH, Via Trevano 131, Lugano 6900, Switzerland
}

\date{Accepted XXX. Received YYY; in original form ZZZ}

\pubyear{2015}
\begin{document}
\label{firstpage}
\pagerange{\pageref{firstpage}--\pageref{lastpage}}
\maketitle

\begin{abstract}

Virtual Observatories (VOs) are online hubs of scientific knowledge. They encompass a collection of platforms dedicated to the storage and dissemination of astronomical data, from simple data archives to e-research platforms offering advanced tools for data exploration and analysis. Whilst the more mature platforms within VOs primarily serve the observational community, there are also services fulfilling a similar role for theoretical data. Scientific visualization can be an effective tool for analysis and exploration of datasets made accessible through web platforms for theoretical data, which often contain spatial dimensions and properties inherently suitable for visualization via e.g. mock imaging in 2d or volume rendering in 3d. We analyze the current state of 3d visualization for big theoretical astronomical datasets through scientific web portals and virtual observatory services. We discuss some of the challenges for interactive 3d visualization and how it can augment the workflow of users in a virtual observatory context. Finally we showcase a lightweight client-server visualization tool for particle-based datasets allowing quantitative visualization via data filtering, highlighting two example use cases within the Theoretical Astrophysical Observatory.

\end{abstract}




\section{Introduction}
\label{sec:introduction}

Data intensive science, heralded as the fourth paradigm of science, revolves around the use of data intensive computing for scientific discovery. One of the first fields to undergo a paradigm shift towards data intensive science was astronomy. Data generated by observations and theoretical simulations has been continually growing in complexity and size due to technological advances in instrumentation and computing systems, which has caused astronomers to face increasingly difficult processes of data storage and access. These difficulties can be exacerbated by logistical problems relating to long term maintenance, and can create barriers that prevent data sharing and inhibit collaboration. Physical storage and transport, security protocols, and complex file formats are just some of the problems that can prevent researchers from effectively managing their data and lead to an environment that is not conducive for cross-institutional and international collaboration.

It has been commonplace for data-intensive scientists to employ online databases and associated web portals for many years now in an effort to improve data accessibility for the scientific community \citep{Szalayetal2006}. The increased availability of data provided by these services allows scientists to publish original research despite not having been part of the original data production team. One of the first international collaborative efforts to aggregate and share access to data in astronomy resulted in the Virtual Observatory (VO) concept, a collection of services with defined standards for storage and access. A history of the VO is provided by \citet{Djorgovskietal05}. This concept and other efforts are supported by the ubiquity and continuous advancement of modern web technologies, enabling accessible-anywhere web platforms that bring scientists ever closer to their big research datasets. In some cases this extends to e-research platforms with multiple in-built services to support a community of scientists in the process of discovery, from data search, subset, and download, to automated creation of derived datasets and online analysis.

Some notable examples include the extensive services incorporated into the Strasbourg Astronomical Data Center (CDS)\footnote{http://cdsweb.u-strasbg.fr/}, and projects related to the International Virtual Observatory Alliance (IVOA)\footnote{http://www.ivoa.net/} such as the German Astrophysical Virtual Observatory (GAVO)\footnote{http://www.g-vo.org/pmwiki/Main/HomePage}, and the All-Sky Virtual Observatory (ASVO)\footnote{http://www.asvo.org.au/}. These are on-line umbrella services that have historically hosted very large repositories of observational images and surveys in standardized formats, alongside web-based tools to assist in finding, identifying, and extracting data from the archives, for example Vizier \citep{Vizier00} and Aladin Lite hosted at the CDS. 

In the case of theoretical data, such as large cosmological simulation outputs, a public data release will often link to an online database where users can download some or all of the data. For example, the Millennium Simulation Data Archive \citep{Lemsonetal06} is hosted through GAVO and accessible via SQL query, and the outputs of the EAGLE simulation \citep{Mcalpineetal16} are hosted and query-able in a similar manner at the Institute for Computational Cosmology of Durham University. While this is undoubtedly beneficial for scientists, access to the raw data is not the only solution that can be offered by data portals, and is sometimes the least convenient. The raw size of data releases means a potentially non-trivial process of identifying and extracting useful data can become prohibitive without dedicated computing resources.

There is now emerging a more advanced type of web portal for theory data that not only provides access to the raw outputs of simulation, but includes tools to assist the astronomer in exploring the data as well as creating or accessing derived datasets that can be more directly useful for their science case. An early example of this type of repository is the Mock Map facility \citep{Blaizotetal05}, an online tool that allows users to access mock galaxy catalogues generated via the GALICS semi-analytic model \citep{Blaizotetal04}. More recent examples (see review in Section \ref{sec:webobservatories}) may also have powerful hardware back-ends  such as High Performance Computing (HPC) systems to support more complex tools for selection, customization and generation of derived data, enabling the processing of large and complex datasets through sophisticated on-line services. 

Scientific visualization has been applied in many different contexts in an effort to support the knowledge discovery process, as can be seen in the comprehensive reviews on visualization techniques for general physical sciences, \citep{Lipsaetal12}, and specifically for astronomy \citep{hassan11}. It is already commonplace to exploit visualization in virtual observatories; observational data is represented with all-sky and per-object image compositions and theoretical data with virtual observations and 3d visualization images. Furthermore, interactive visualization is commonly used in the case of 2d observational images. For example, while searching for and extracting observational data objects, a user typically visualizes and navigates images in their web browser using an interactive 2d interface such as the Aladin Lite Sky Atlas \citep{Aladinlite14}.

Ideally, a user should be able to follow a similar interactive navigation process to access theoretical data. However, data from computer simulations is frequently expressed in three spatial dimensions, and more naturally explored via interactive 3d scientific visualization. This is a more difficult service for a web portal to provide because on-demand 3d visualization can be computationally expensive, especially so in this context as the size and complexity of theoretical data can easily exceed the capabilities of modern web browsers, and is continually growing. 

Astrophysical simulations can have extremely high computational and memory requirements, requiring large scale solutions provided by HPC facilities. For example, the raw data of the flagship Millennium run consisted of over 10 billion particles and was executed on the Max Planck Societies principle HPC system for over a month, with the final output requiring roughly 25 TB of storage. These types of data products are far too large to fit in the memory of a personal workstation and require large scale computing resources to host and process. Due to the fact subsequent processing and analysis requires HPC as well, it is a natural consequence that web portals for these data products are also being supported by HPC systems.

The key topic of this paper is to demonstrate how interactive 3d visualization can support users of virtual observatory interfaces in accessing, filtering, exploring, and extracting theoretical data. We consider the potential to further utilize HPC resources to provide 3d visualization directly within web portals, as an additional service to support the scientist in management and analysis of large theory datasets. 3d visualization can be exploited to improve the efficiency of accessing data, and particularly interactive visualization can provide real-time visual feedback to further support the exploration and extraction process. We aim to ensure that larger datasets and demanding processing tasks remain within the web portal, while the user only downloads and processes the minimum amount of data they require. Those larger datasets and tasks can then further rely on the support of HPC resources to ensure we have the performance necessary to visualize and interact in real-time.

To this end we review the current state of the art in 3d visualization within web-based astronomy data portals, and present a approach based on the open source visualization software Splotch \citep{Dolagetal08} with discussion of its application for data selection and exploration. Splotch, discussed further in Section \ref{sec:splotch}, is a high performance code designed for HPC systems, with recent developments including a client server model with web interface and data filtering features. We exploit these developments to demonstrate the utility of 3d visualization within theoretical astronomical data portals, with examples in the context of the Theoretical Astrophysical Observatory (TAO) \citep{Bernyketal16}.

The paper is structured as follows: Section \ref{sec:webportal3dviz} reviews current online web-portals for theoretical astronomical data with a focus on visualization capabilities, and presents a brief discussion of the challenges inherent in 3d web-based visualization in this context. Section \ref{sec:splotch} details a novel approach for interactive web-based visualization we have developed for astrophysical observatories building upon an existing visualization software, with Section \ref{sec:quantitativeviz} detailing the quantitative visualization process enabled. Contextual application, within specific scientific use cases, is shown in Section \ref{sec:appsincontext}. Discussion of the work in Section \ref{sec:discussion} compares with related tools and highlights longer term goals, with final conclusions presented in Section \ref{sec:conclusions}.

\section{Visualization for Theoretical Web Portals}
\label{sec:webportal3dviz}

There is a new type of theoretical astronomical web portal emerging, exploiting advances in modern web technologies and powered by high performance back ends. In this section we focus on the visualization capabilities of those characterized by hosting large theoretical data for public access with the support of high performance computing resources.

\subsection{The Web Observatories}
\label{sec:webobservatories}

Current theoretical astronomical portals have visualization capabilities as summarized in Table~\ref{tab:vizportals}. We take note of three types of visualization capability:

\begin{description}
\item \textbf{2d:} Two dimensional plotting capabilities such as histograms, line graphs, scatterplots.
\item \textbf{3d:} Three dimensional plotting capabilities such as maps and 3d renderings. 
\item \textbf{VObs:} Virtual observations generated by scientific mock imaging packages.
\end{description}

For each type of visualization we note the availability type, either \emph{user-generated} or \emph{pre-computed}, which distinguishes respectively whether the user can generate their own new visualizations or simply browse a selection of existing visualizations. 

We further note the type of interaction available: \emph{static} in the case of simple images; \emph{2d interaction} in the case of Zoomify-like\footnote{http://www.zoomify.com/} navigable images; and \emph{3d interaction} for 3d rotate and zoom navigation such as found in traditional interactive 3d visualization packages.

\begin{table*}
	\centering
	\caption{Visualization in Theoretical Astronomy Portals}
	\label{tab:vizportals}
	\begin{tabular}{lccr} 
		\hline
		\textbf{Portal} & \textbf{Visualization} & \textbf{Availability} & \textbf{Interaction}\\
		\hline
        \citep{Overzieretal13} & VObs 	& pre-computed   & 2d interaction \\ \hline
		\citep{Chardetal14}    & 2d 	& user-generated & static         \\
          					   & VObs 	& user-generated & static         \\ 
                               & 3d 	& user-generated & 3d interaction \\ \hline
		\citep{Nelsonetal15}   & VObs 	& pre-computed   & static         \\ 
          					   & 3d 	& pre-computed   & 2d interaction \\ \hline
		\citep{Bernyketal16}   & VObs 	& user-generated & static         \\ \hline 
        \citep{Ragagnin16}     & 2d 	& user-generated & static		  \\
        					   & VObs 	& user-generated & static         \\ 
        				       & 3d 	& pre-computed   & 2d interaction \\
		\hline
	\end{tabular}
\end{table*}

The Millenium Run and later simulations are housed in the Millenium Run Database, an SQL-queryable database for the original data outputs of the simulations \citep{Lemsonetal06}. This is combined with the Millenium Run Observatory \citep{Overzieretal13} hosting lightcone catalogues with virtual observations in many different configurations and other associated data, all of which can be traced back to the data products hosted in the Millenium Run Database.

PDACS \citep{Chardetal14} is a web portal and science gateway built upon the GALAXY workflow engine for life sciences research\footnote{https://galaxyproject.org/}, re-purposing the engine as a platform for data access and analysis for a suite of cosmological data products forming the Coyote universe \citep{Lawrenceetal10}. The platform is supported by NERSCs computing infrastructure and ANLs scientific cloud. It incorporates a variety of data analysis services, including ParaviewWeb (discussed further in Section \ref{sec:webvizastro}) integration that enables interactive 3d visualization\citep{Maddurietal15}.

The Illustris project \citep{Vogelsberger14} released a web portal\footnote{http://www.illustris-project.org/} along with their public data release \citep{Nelsonetal15}  which hosts query-able data produced by the suite of Illustris simulations along with various tools for data searching and exploration. The site is supported by a cluster back-end and includes \emph{The Explorer} for 2D exploration of Illustris visualizations, along with the \emph{Galaxy Observatory} which provides mock images of galaxies at z=0 where stellar mocks have been built. While visualizations are currently pre-computed, the authors indicate that future promising directions include providing tools and resources for users to compute their own arbitrary results from the hosted data.

\citet{Bernyketal16} describe the Theoretical Astrophysical Observatory (TAO). TAO is an online virtual observatory providing access to mock extragalactic survey data generated by running complex semi-analytic galaxy formation models on the output of large N-body cosmological simulations. TAO provides a variety of science modules for users to generate and extract data useful for their specific science case, supported by a supercomputing system as a back-end. In terms of visualization, users can apply a science module to generate custom mock image observations. TAO is an example-case and motivation of the presented effort toward incorporation of interactive 3d visualization to virtual observatories, and discussed further in Section \ref{sec:appsincontext}.

The Cosmological Web Portal \citep{Ragagnin16} is an online web service for hydrodynamical, cosmological simulations. Supported computationally by LRZ and C2PAP, the portal provides services to browse and subset Magneticum data\footnote{http://www.magneticum.org/} as well as generate various 2d maps, virtual observations and spectra. Visualizations in 3d are pre-computed via the Splotch software. These services create jobs on the underlying HPC system in order to satisfy user requests, and so allow users to not only explore existing raw and derived data, but generate further results to suit their own needs.

As can be seen from Table~\ref{tab:vizportals}, in general it is not common to include interactive 3d visualization for astronomical web portals. The exception, \citet{Chardetal14}, is the most advanced platform, built upon an existing workflow engine and customized for cosmological workflows. A key difference between PDACS and the other portals presented is that it is a workflow engine for cosmology, rather than a public access repository. For example, in order to use PDACS a user must have an associated account with computing allocation. 

A strong factor in the lack of 3D vizualization for web portals is the computational expense. Large theoretical datasets require HPC resources for visualization, however achieving interactivity is not trivial on HPC systems due to both the computational expense and factors such as scheduling and integration of web and HPC applications. Technical challenges are further discussed in Section \ref{sec:webvizastro}.

\subsection{Visualization as part of the Data Access Workflow}
\label{sec:dataaccessworkflow}

\begin{figure}
 	\includegraphics[width=\columnwidth]{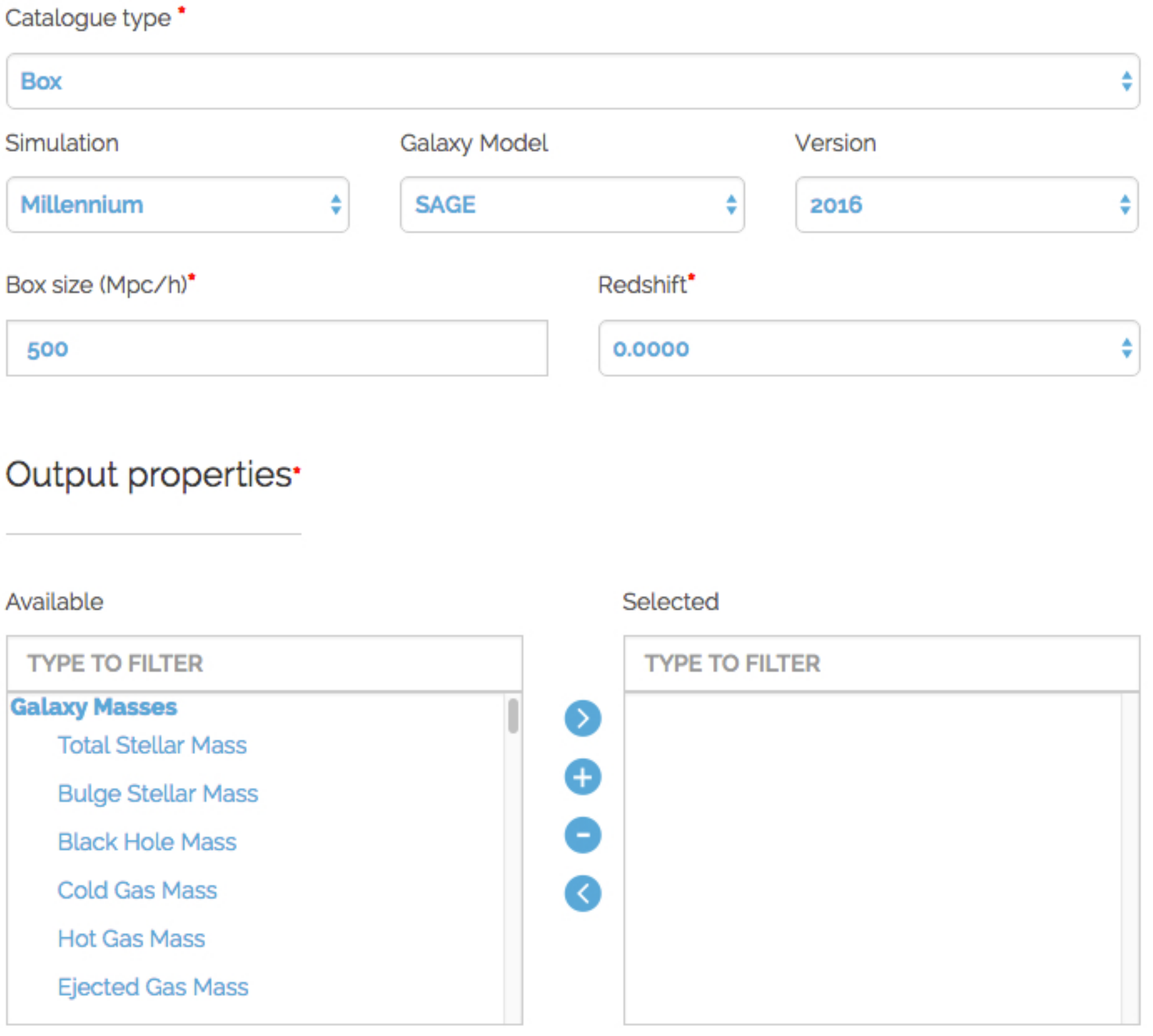}
     \caption{Example section of the interface for specifying and extracting data from the Theoretical Astrophysical Observatory.}
     \label{fig:taowebform}
 \end{figure}

\begin{figure}
 	\includegraphics[width=\columnwidth]{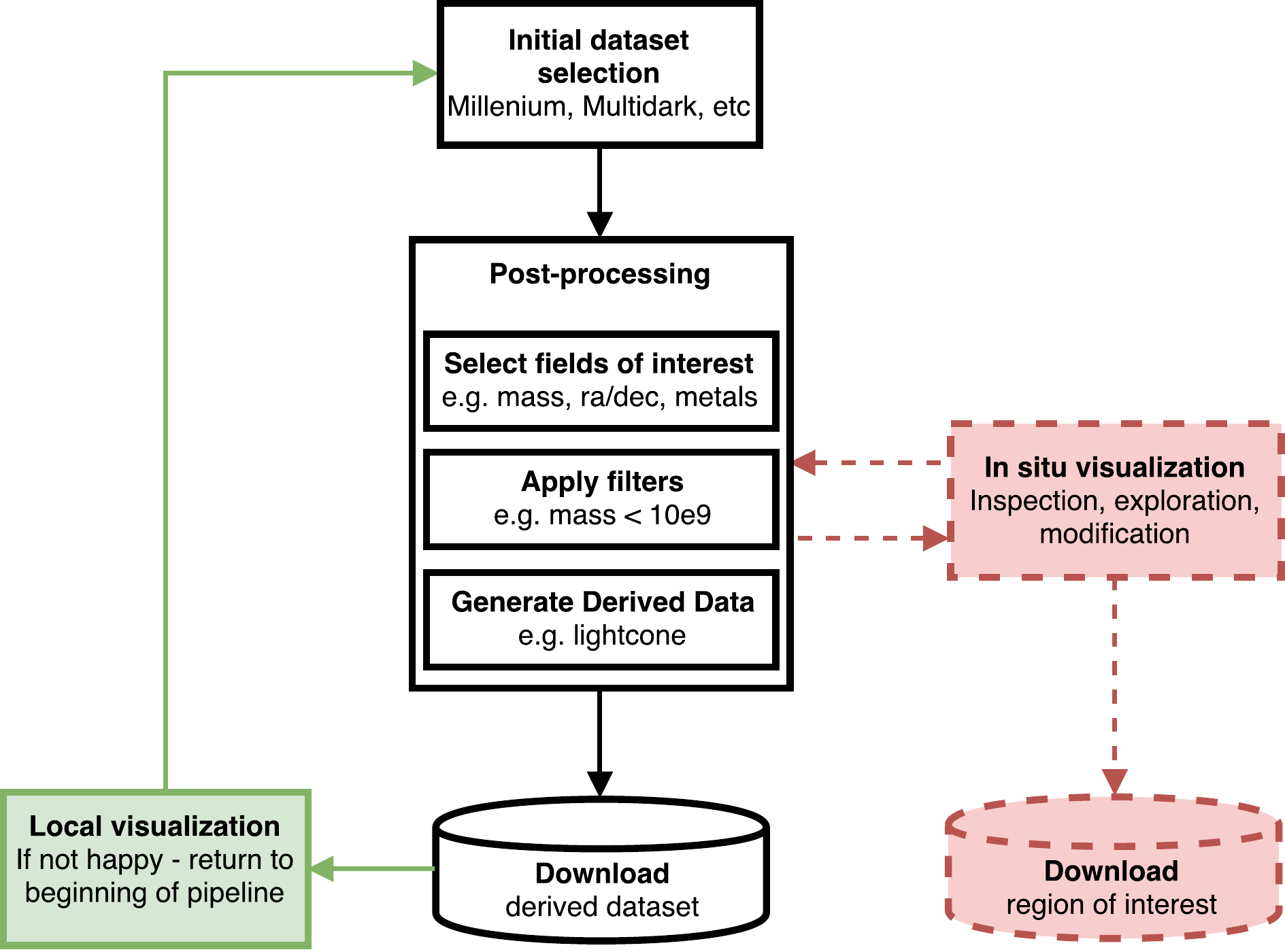}
     \caption{The process for a user to access theory data from a web portal. The feedback loop in green highlights the current process for data inspection, while the feedback loop in red shows the proposed process.}
     \label{fig:dataaccessworkflow}
 \end{figure}

Visualization is an essential factor in effective data exploration, and we explain here how it can also support the data access workflow. The workflow in Figure \ref{fig:dataaccessworkflow} is a minimal representation of the steps a scientist, or user, would take to access data within a web platform. The user will start by defining an initial dataset, and apply post-processing to build a derived dataset. This may include an extraction process in which the user applies filters to constrain data properties and extract a specific data subset (see example interface from the TAO platform in Figure \ref{fig:taowebform}), as well as applying transformations via science modules made available to the user. Finally they may download the resultant dataset(s).

In current web repositories without on-demand analysis, the user must download their data before performing some initial data processing to confirm it is as required. This may be done with a variety of analysis techniques, often including local visualization. However, if after some analysis and exploration it is clear that the data is not exactly as required, the user may need to return to the web portal and begin their process again. This feedback loop represents a trial and error process, an unnecessary barrier that can be time consuming and interrupt the users workflow. It is highlighted in Figure \ref{fig:dataaccessworkflow} as the solid-line green box on the left side of the figure. 

Conversely, the dotted red box in Figure \ref{fig:dataaccessworkflow} represents the inclusion of in-situ and on-demand analysis techniques. In this case, in-situ refers to the preliminary analysis being performed within the portal where the data is situated, while on-demand refers to the use of scheduled HPC resources to fulfill analysis task requirements. In-situ and on-demand visualization allows the user a quick view of their dataset during the filtering and transformation process, helping to identify areas of interest or visible errors before downloading the data. 

Furthermore, an interactive visualization tool that includes some built-in means of manipulation and filtering can enhance this process by allowing the feedback loop between visual examination and filter application to take place interactively. This enables a more intuitive exploration and filtering process where the user can add and remove experimental filters as necessary with real-time visual feedback (discussed further in Section ~\ref{sec:quantitativeviz}). The ultimate goal is to provide the scientist with a clearer understanding of their data before leaving the portal to apply their own methods.

\subsection{Technical Challenges for Web Visualization in Astronomy}
\label{sec:webvizastro}

In order to exploit scientific visualization on the web, visualization tools must be able to communicate with the ecosystem of software that makes up modern web applications. For example, current browsers come equipped with powerful engines for Javascript, which has become the de-facto language of the web. As such, the various approaches for 3d rendering in a web browser are predominantly Javascript libraries based on the WebGL graphics API (for a thorough review of these technologies the reader is referred to \citet{Evans20143DGO}). The most notable web-only approach to 3d scientific visualization is the currently ongoing effort to port a subset of the Visualization Toolkit library\footnote{https://www.vtk.org/} to Javascript as vtk.js\footnote{https://kitware.github.io/vtk-js/}. Further approaches can be seen in the recent state of the art review of web-based visualization techniques \citep{Mwalongo16}.

Whilst web-only approaches to visualization would be ideal, two of the most challenging technical aspects of 3d visualization for astronomy are the computational and memory requirements. Theoretical astronomical datasets are often orders of magnitude larger than one could store and process on a home desktop or laptop. For example, the semi analytic galaxies currently stored on TAO\footnote{https://wiki.asvo.org.au/display/TAOC/Available+Data+Sets} are built from N-Body simulations ranging up to tens of billions of particles, while at the extreme scale modern day simulations are advancing even further with a recent run of the PKDGRAV cosmological code exploiting two trillion particles and requiring roughly 124 terabytes of storage \citep{Potteretal17}. With current web-only visualization tools it is not possible to visualize data of this size interactively in 3d solely via web browser. Instead, as seen in \citet{Mwalongo16}, it is possible to perform remote 3d rendering via a powerful back-end computing system connected to a web based client in a client-server architecture. 

The concept of performing large data processing as physically close as possible to the storage is coined \emph{moving computing to the data}. Transferring large amounts of data to the client for rendering is often impractical, if not simply impossible, due to data size and network transport. However, sending compressed images is often a much easier task and it is well within the capabilities of modern browsers to receive and display these at high speed. In this context it is common to use a client-server architecture that performs compute tasks on a server close to the data and streams images or video to a client, reducing large data transfers over wide-area networks such as the internet. 

There are two notable solutions for client-server web based visualization for astronomy. Firstly, the well-known general purpose scientific visualization tool Paraview \citep{Ahrensetal05} and associated web-toolkit ParaviewWeb\footnote{https://github.com/Kitware/paraviewweb}. The web-toolkit allows developers to integrate a web-client for the Paraview server into their application. While the web framework is relatively recent it is the most advanced tool currently for web based scientific visualization, with options from web-only visualization to a full client-server mode with a Paraview or VTK \citep{Schroederetal06} backend. 

VisIVO is a suite of visualization software targeting the virtual observatory environment \citep{Comparatoetal07} \citep{Beccianietal10}. The suite includes a client-server tool supporting grid computing systems, which can integrate with a web-based science gateway tool built around grid-computing middle-ware workflow tools\citep{Sciaccaetal13}. The underlying viewer can exploit both VTK and Splotch as rendering engines. VisIVO is currently under development to extend capability for visual analytics \citep{Sciaccaetal15}, and is currently the most VO-compliant scientific visualization tool.

A more recent concept that has gathered momentum is the use of cloud technologies for high performance processing, including visualization. NVIDIA offer a GPU Cloud service that can provide high performance visualization capabilities on cloud based GPU hardware with minimum setup and installation\footnote{https://www.nvidia.com/en-us/gpu-cloud/hpc-visualization-containers/}. There is not yet much progress toward utilizing a service such as this for visualization in a virtual observatory setting. A key hurdle to this approach is the location of the source data, usually already stored in a HPC center. To use cloud based visualization services the web platform must also store the data within the cloud, which can be costly. However, as the technology matures and cost is reduced future efforts may be able to exploit cloud based technologies for this type of high performance visualization directly within a web platform. 

\section{Advanced Particle Visualization for Web}
\label{sec:splotch}

\subsection{Splotch}
\label{sec:thecode}
\begin{figure*}
   	\includegraphics[width=\textwidth]{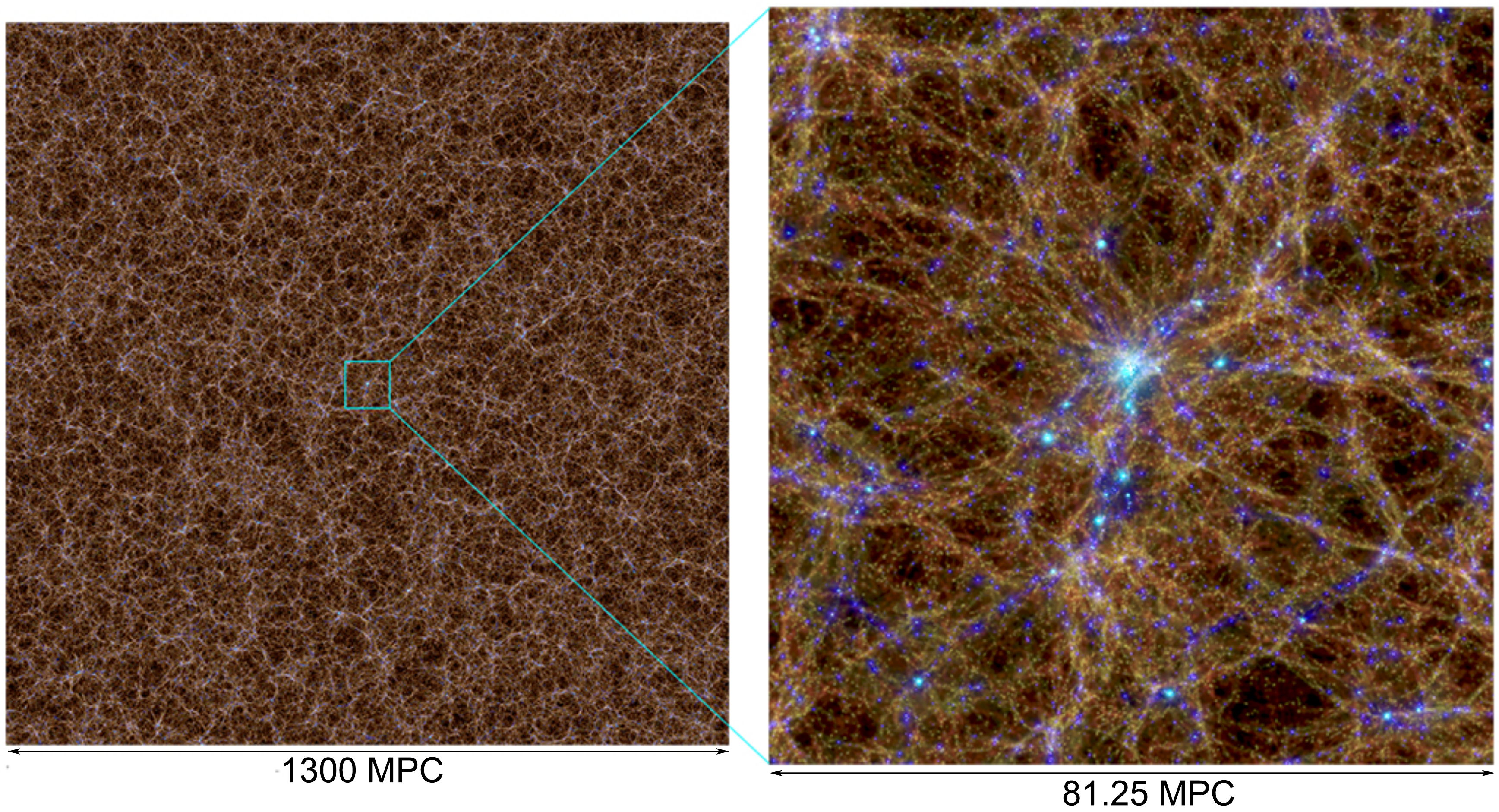}
    \caption{A flagship Splotch visualization. The image depicts the Magneticum simulation, Box/mr0, consisting of 1.86x10\textsuperscript{11} particles \citep{Dolagetal15}. The full size image (left) is rendered in 16000\textsuperscript{2} pixels and shown alongside a 16x zoom-in image of the box center (right).}
     \label{fig:Magneticum}
 \end{figure*}

\begin{figure}
 	\includegraphics[width=\columnwidth]{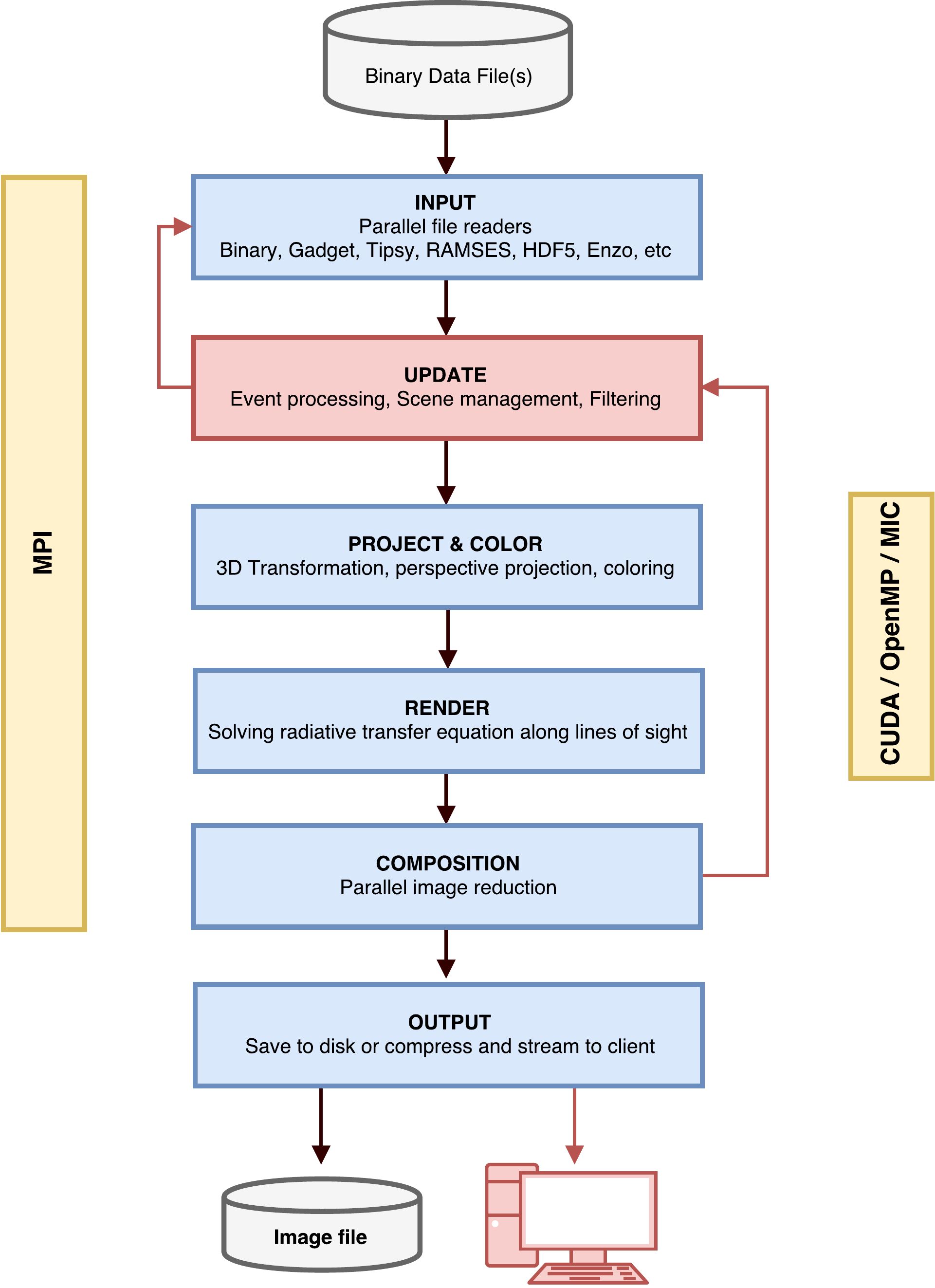}
     \caption{The software structure of the Splotch code, describing the main algorithm stages, various parallel models and a non-exhaustive list of available file readers. The sections highlighted in red are key new additions to support a client-server interactive model, with significant re-design and optimization throughout the pipeline to support interactivity and data filtering. }
     \label{fig:SplotchStructure}
 \end{figure}

Splotch is an open source scientific visualization tool designed for high performance computing environments. Written in C++, with minimal external dependencies, Splotch is implemented with a variety of parallel models including hybrid OpenMP and MPI for x86\_64 environments such as Xeon and Xeon Phi CPUs, as well as CUDA \citep{nvidiacuda} for GPU based systems \citep{RiviEtAl13} \citep{Jinatal10}. The lack of dependencies and inclusion of multiple parallel models means Splotch can be built to exploit almost any type of system, from a standalone desktop system with or without a GPU to large heterogeneous HPC architectures. The key strengths of the Splotch software are: support of very large particle based datasets and diverse HPC systems (Splotch has successfully visualized 1.86x10\textsuperscript{11} particles, pictured in Figure \ref{fig:Magneticum}), and fast, high quality volume rendering of particle datasets such as those resulting from SPH simulations. We choose to use Splotch as the visualization engine for this work due to these key strengths. Figure \ref{fig:SplotchStructure} illustrates the structure of Splotch, with various readers for standard binary files and common astrophysical file formats included. 

Recent developments in Splotch include moving from a traditional batch HPC software approach to an interactive client-server model. The traditional batch approach, i.e. where the user submits a job script and waits for the job to be scheduled and run producing a collection of static images, is acceptable for pre-rendered visualizations and movies, however for interactive visualization it is necessary to have access to the resources during job execution. Full technical details of this approach are out of the scope of this paper and will be presented (in preparation), however a brief overview is given here for context.

 \begin{figure}
 	\includegraphics[width=\columnwidth]{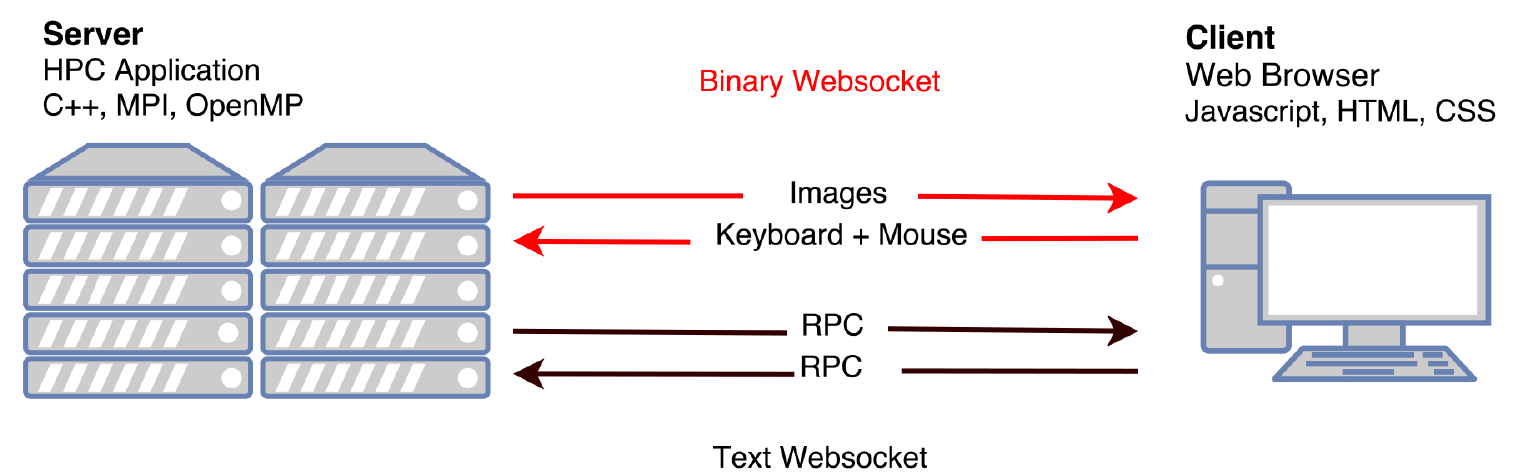}
     \caption{Splotch: Client server framework with Websocket based communication.}
     \label{fig:clientserver}
 \end{figure}

The new client-server Splotch consists of a rendering server that can run on a remote HPC system, which streams visualization results to a web client running on a user machine. The data to be visualized is stored on the high performance back end, and only compressed images are sent to the client. The server receives input asynchronously from the client in the form of interaction events such as user connection, key and mouse events, and remote procedure calls (RPCs), all processed in order of receipt during an event handling stage at the beginning of each frame. The scene is updated accordingly, and results are generated, compressed, and streamed to the web client as JPEG images. 

The client is a standalone web application written entirely in Javascript, embedded in a HTML web page with CSS formatting. A default interface consisting of an image window and debugging console is defined on the client while any further user interface elements are generated on-the-fly from an interface descriptor received from the server. The default interface is built using a quick and simple UI library, dat.gui\footnote{https://github.com/dataarts/dat.gui}, however the design is such that implementors can and would override this using any other user interface library. The key feature of this client is that it is \emph{thin}, there is minimal processing happening on the user machine and as much as possible is relegated to the high performance server. 

Communication is handled via the WebSocket protocol \citep{RFC6455}, and illustrated in Figure \ref{fig:clientserver}. A primary Websocket provides two-way binary streaming for images and high frequency events (mouse and keyboard), and a secondary WebSocket provides text based JSON\footnote{JSON Standard: https://tools.ietf.org/html/rfc7159} request-response message passing (for RPC calls). The rendering server and interaction client both act as simultaneous RPC client-servers, i.e. on either end of communication RPC requests can be both sent and received. These are formatted in a simple and portable manner through the JSON-RPC protocol\footnote{http://www.jsonrpc.org/specification}. The structure of the Splotch client-server implementation allows a thin web-client to be embedded in a web service, connecting a web-based user (e.g. the TAO service) to an interactive visualization environment. 

Alternatively the user can run both client and server on their laptop or a desktop with or without GPU for local visualization of smaller datasets that do not necessitate a high performance system.

\subsection{Data Filtering for Quantitative Visualization}
\label{sec:quantitativeviz}

Quantitative visualization allows the user to directly tie the numerics of their data to the visual output of a visualization, inferring quantitative information from visual presentation \citep{Peskinetal91}. This is achieved by providing ways to perform quantitative data processing as part of the visual interaction process. We approach this in Splotch by including a data filtering menu, which allows the user to perform a variety of quantitative tasks on their dataset. Currently supported are three types of filter: clipping, arithmetic, and combinatory.

\textbf{Clipping} filters allow the user to subsample their data by inputting ranges of any existing field in the data (including those not currently being used in the visualization). An example of this could be clipping a lightcone geometrically in Right Ascension and Declination, followed by applying a mass cutoff to show only very high mass galaxies.

\textbf{Arithmetic} filters provide a way of scaling or otherwise arithmetically modifying any existing field of the data. An example of this could be applying a logarithmic filter to the visualized quantity to better distinguish fields across a high dynamic range, or scaling data quantities for use e.g. in combinatory filters.

\textbf{Combinatory} filters allow the combination of multiple fields of the data to create derived fields for visualization. A user can provide data fields as left and right hand sides of a statement, and pick a combinatory function to apply such as adding or subtracting. This can, for example, be used to subtract colour bands from one another in a galaxy survey, the differences of which can point to patterns in the data such as concentrations of old vs young stars or be used as a proxy for redshift. 

Filters can be stacked on top of one another which allows a user to interactively build e.g. advanced selection cuts for a galaxy catalogue. An example of the potential applications of combinatory and arithmetic cuts can be seen in \citet{Eisensteinetal01}, where multiple colour bands are combined and various scale factors applied to create derived fields that are used to select luminous red galaxy targets from the Sloan Digital Sky Survey. The process of quantitative visualization through data filtering allows a user to interactively experiment with selection cuts such as these, with interactive visual feedback from their data to aid the scientific process.

Once satisfied with the results of an interactive filtering process, the user can opt to store the filtered dataset to a specified location on the server. Embedded in a web service, this extends to a direct download of the dataset via HTTP. Additionally, a metadata file is stored containing a JSON representation of the server state and applied filters, which can be used to reload a desired scene and filter configuration.

\section{Visualization In the Context of TAO}
\label{sec:appsincontext}

In this section, we present several use cases applying our tool to TAO data, demonstrating the current capability to build up advanced filters for target selections and to extract useful subsets from a larger scientific dataset. This approach could in future be similarly adopted by other theoretical web platforms.

TAO is a web platform hosting data from multiple cosmological dark matter numerical simulations and galaxy formation models. TAO includes various science modules for users to apply to their data, this includes extracting a light cone with parameters that match particular survey geometry, or generating custom mock images through SkyMaker \citep{BertinEtal09}. TAO is an example-case and motivation of the presented effort toward incorporation of interactive 3d visualization to virtual observatories, and discussed further in Section \ref{sec:appsincontext}. A node of the ASVO project\footnote{http://www.asvo.org.au/}, TAO is supported by the Center for Astrophysics and Supercomputing (CAS) at Swinburne University of Technology, Melbourne, Australia. In addition to support and administration, CAS provides use of the Swinburne HPC cluster as a high performance back-end to support the cloud-based platform. As a web platform for theoretical data supported by HPC resources, and built in a modular way to enable integration of new science modules, TAO is an ideal example platform to demonstrate the utility of 3d visualization in this context.

Users are able to access the database through a point and click web-client, and can exploit various science modules for tasks such as constructing observer light-cones, spectral distributions and virtual observations. TAO has been used as a source of data for many science users (e.g. Section \ref{sec:evironmentexploration}), and provides a vital service to the Astronomical community. Starting from existing scenarios utilizing data from the TAO database, we replicate some of the early steps taken in the scientific process to demonstrate the utility of an interactive 3d web visualization tool in this context.

Framerates reported in this section are recorded with the Splotch visualization server running on Swan, a Cray XC50 located at the Cray computing facility in Wisconsin, USA. Each node consists of 2 Broadwell 22-core Xeon CPUs clocked at 2.2Ghz. The web client is running on a Macbook Pro (early 2013 model) with 2.7 GHz Intel Core i7, and Mozilla Firefox 57.0.4, at the University of Portsmouth in the UK. 

\subsection{Exploring the environments of recently merged galaxies}
\label{sec:evironmentexploration}

One example of the power of interactive data visualisation for a service like TAO is as a tool to find the large-scale environments of recently merged galaxies. Galaxy-galaxy collisions are common and can have dramatic effects on the resulting galaxy's properties. Just where and when these occur is a matter of active study by the community. After creating a mock survey universe in TAO, additional visualisation can be performed using clipping, arithmetic and combinatory filters allowing the user to draw out recently merged systems (or even those about to merge) so that their properties can be studied and cross-correlated with the broader environments in which they live (void, filament, group or cluster). Such information can then be used for mock imaging or for additional processing once downloaded locally, and directly compared with equivalent observational data sets.

\citet{Pennyetal15} sub-sampled galaxies from the Galaxies and Mass Assembly survey \citep{Driveretal09} to study bright galaxies found in voids. One question posed by the authors is whether the stellar mass of these galaxies is primarily attained through star formation or mergers. The study made use of theoretical data in the form of a galaxy catalogue produced by the semi-analytic galaxy model SAGE \citep{Crotonetal16} run upon the Millennium N-body simulation, which includes information unavailable in observational data such as the time since last major merger. The authors extracted these galaxies from the TAO archive via a series of filters in the web interface, followed by further neighbour-based local environment filtering, and presented visualization comparisons (e.g. Figure 12 of \citet{Pennyetal15}).

Investigating the properties of galaxies in cosmic voids is a key science goal. The initial dataset is a 250 h\textsuperscript{-1} Mpc x 250 h\textsuperscript{-1} Mpc x 20 h\textsuperscript{-1} Mpc subset of the galaxy catalog. In many contexts a random selection here is fine, however it is also possible to tune the location of the sample to increase the number of voids and therefore targets for studying. Visualization can allow the user to undertake a visual process of sampling to identify relevant sections of the box, in order to increase the potential number of useful galaxies. 

To illustrate this, we take a selection of galaxies starting with the full SAGE dataset (500 h\textsuperscript{-1} Mpc\textsuperscript{3}) at \emph{z=0}, extracting 3d galaxy positions, \emph{Total stellar mass} and \emph{Time of last major merger}. This consists of 15619131 galaxies. 

\begin{figure*}
 	\includegraphics[width=0.8\textwidth]{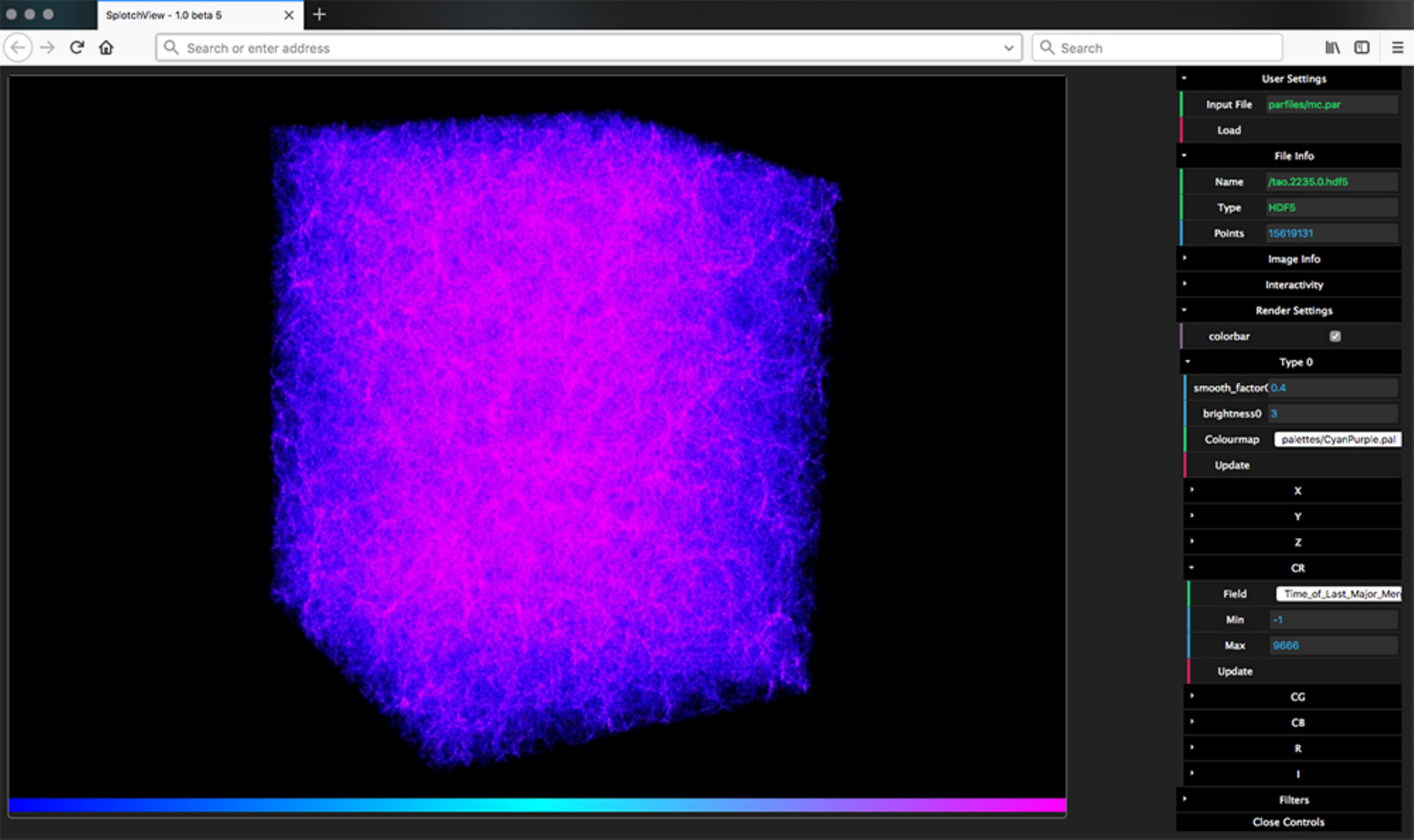}
     \caption{Splotch: the full SAGE dataset at \emph{z=0}, 500 h\textsuperscript{-1} Mpc x 500 h\textsuperscript{-1} Mpc x 500 h\textsuperscript{-1} Mpc, coloured by \emph{Time of last major merger}. The server is running on a HPC cluster, while the client is on a local laptop. Mouse/keyboard controls allow the user to rotate and zoom, while the interface to the right allows to change colour palettes, modify visualized quantities and more.}
     \label{fig:RecentMergers0}
     \includegraphics[width=0.8\textwidth]{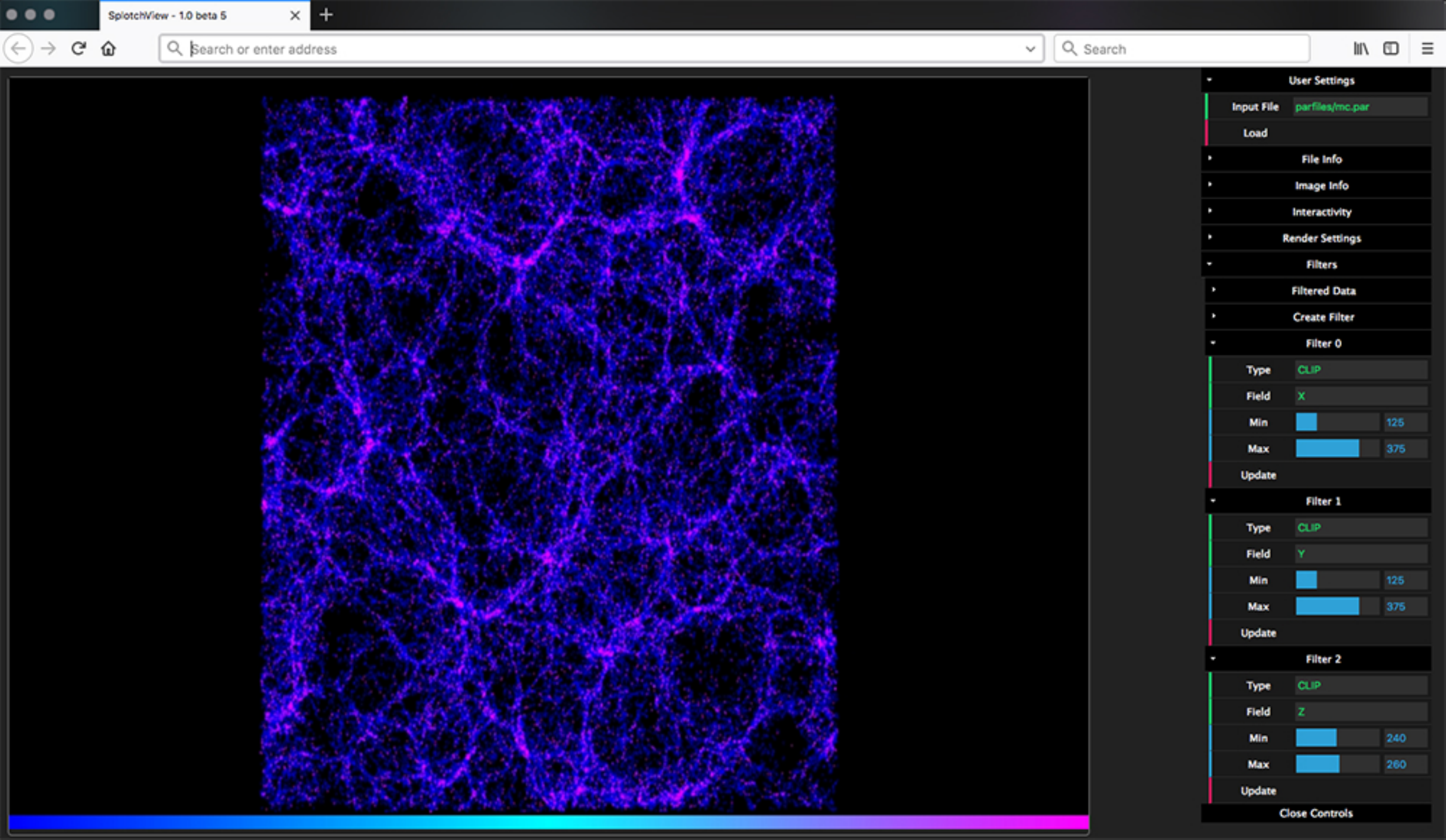}
    \caption{Splotch: clipping filters applied to the data of Figure \ref{fig:RecentMergers0} extract a 250 h\textsuperscript{-1} Mpc x 250 h\textsuperscript{-1} Mpc x 20 h\textsuperscript{-1} Mpc sub-slice. The user zooms in to fill the screen with the filtered dataset. The menu for applying and modifying filters can be seen in the lower right of the figure.}
     \label{fig:RecentMergers0.5}

 \end{figure*}

\begin{figure*}
   	\includegraphics[width=0.8\textwidth]{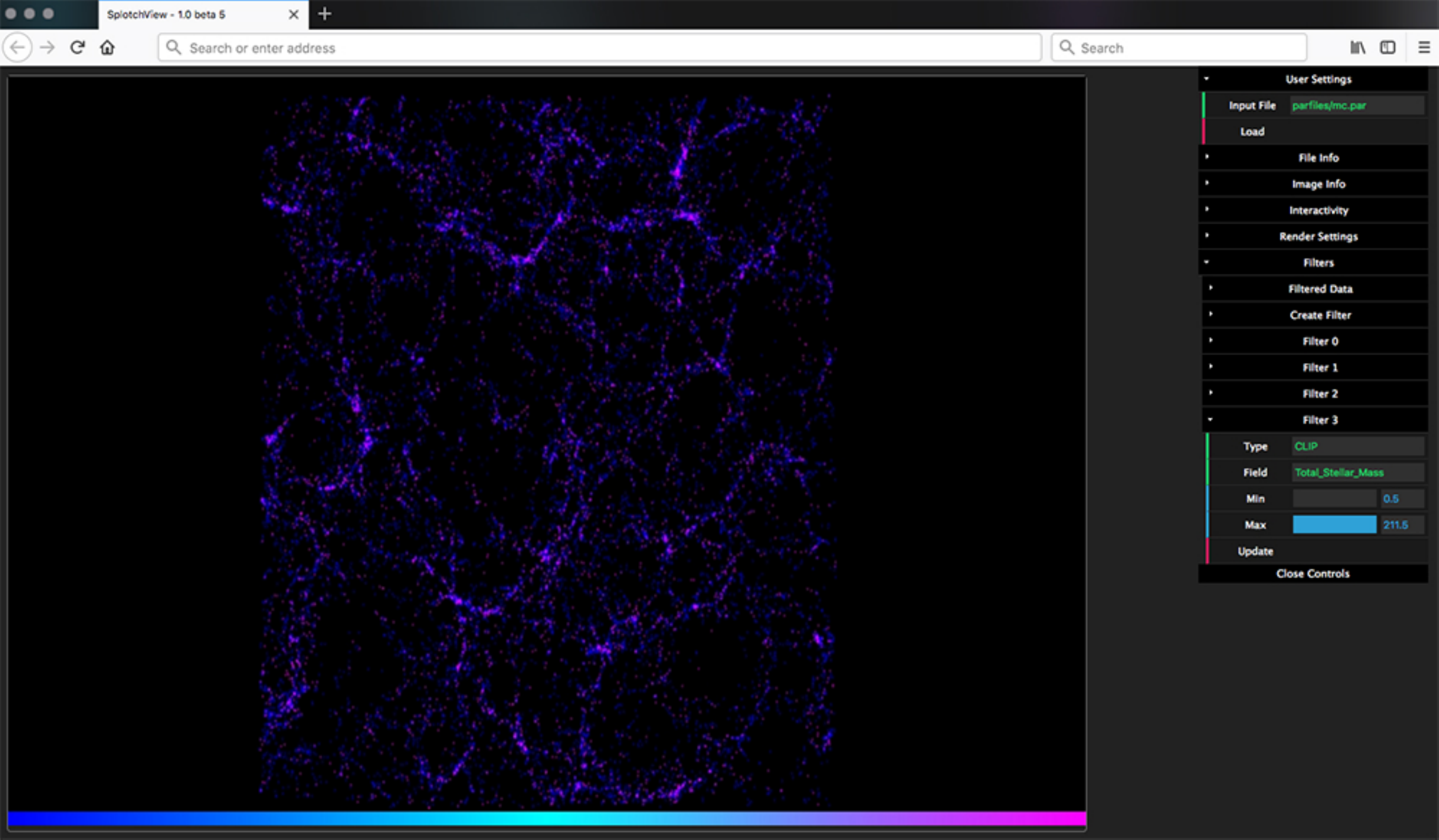}
    \caption{Splotch: further clipping filters applied to the data of Figure \ref{fig:RecentMergers0.5} extract a sub-slice containing only galaxies with stellar mass > 5x10\textsuperscript{9} h\textsuperscript{-1}\(\textup{M}_\odot\).}
     \label{fig:RecentMergers1}
 	\includegraphics[width=0.8\textwidth]{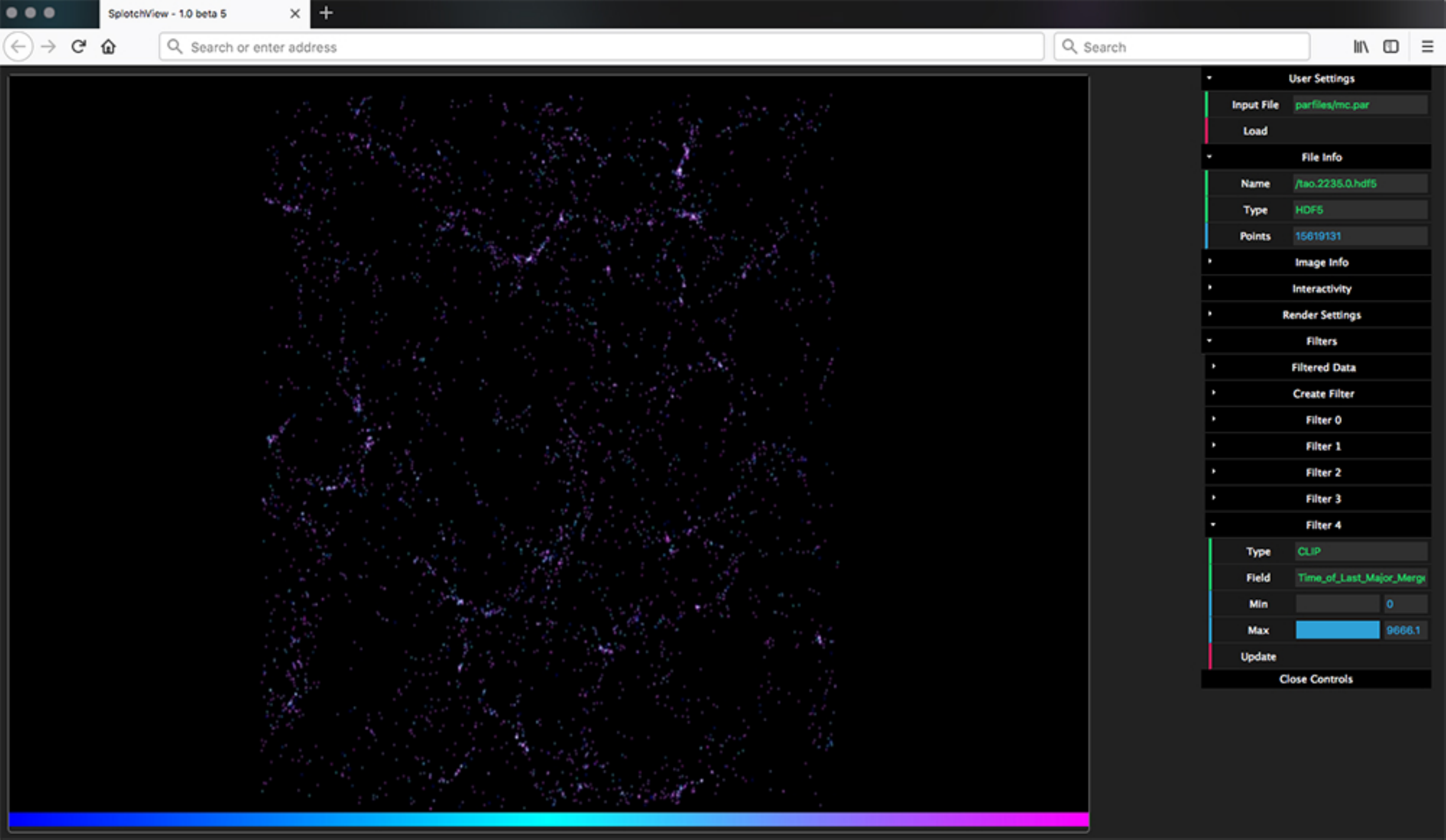}
     \caption{Splotch: Extending the filtering process of Figure \ref{fig:RecentMergers1}, an additional filter is placed on \emph{Time since last major merger} showing only galaxies who have undergone a major merger at some point in their evolutionary history.}
     \label{fig:RecentMergers2}
 \end{figure*}

Using 2 compute nodes of our test system Swan, we are able to visualize the full volume interactively at >10 Frames Per Second (FPS), and begin to visually identify locations to filter remotely from the web browser of a local laptop, demonstrated in Figure \ref{fig:RecentMergers0}. Subsequently, Figures \ref{fig:RecentMergers0.5}, \ref{fig:RecentMergers1}, \ref{fig:RecentMergers2}, and \ref{fig:RecentMergers3}, document the incremental application of filters to match the initial selection criteria of \citet{Pennyetal15}. 

In this use case, the initial selection becomes an interactive process supporting visual input and feedback, with the ability to undo-redo filtering and in real-time explore the environment and history of galaxy mergers before downloading specific subsets for further analysis.

\subsection{Target Selection in the Large Scale Galaxy Distribution}
\label{sec:targetselection}

One common scenario an astronomer faces is to find a specific galaxy population for analysis amongst the wider distribution of galaxies spread across the Universe. For example, astronomers commonly study the physics of massive galaxy formation inside galaxy clusters, which host all kinds of interesting phenomena such as red-and-dead elliptical galaxies and active radio galaxies. Beyond galaxy evolution, such clusters are also important probes of the cosmology of the Universe and can be the focus of highly sophisticated observing campaigns (e.g. eRosita \citep{Merlonietal12}). Using TAO one can build a mock universe that includes many millions of galaxies out to great distances. Currently, to sub-select from this larger population and identify e.g. clusters, the data needs to be first downloaded and processed by the user locally with their own code. Any subsequent imaging within TAO then requires the user to run a new job using this new information, duplicating effort and data. 

\begin{figure*}
 	\includegraphics[width=0.8\textwidth]{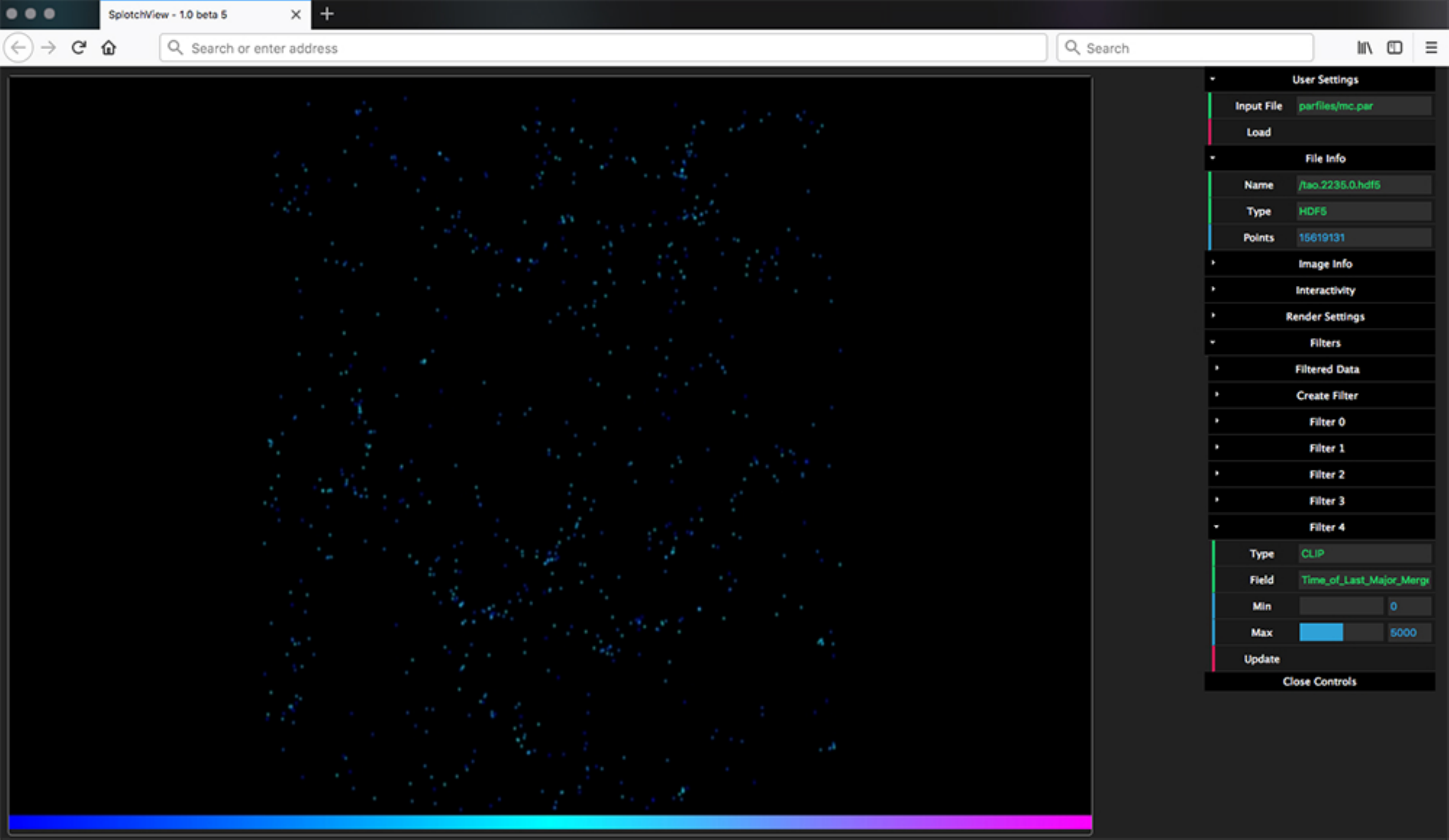}
     \caption{Splotch: Exploring the merger history of galaxies,  the filter on \emph{Time since last major merger} is modified to only include galaxies with a merger in the last 5 Gyr. The filter menu allows this dataset to be saved locally, along with visualization images and interaction history.}
     \label{fig:RecentMergers3}
     
 	\includegraphics[width=0.8\textwidth]{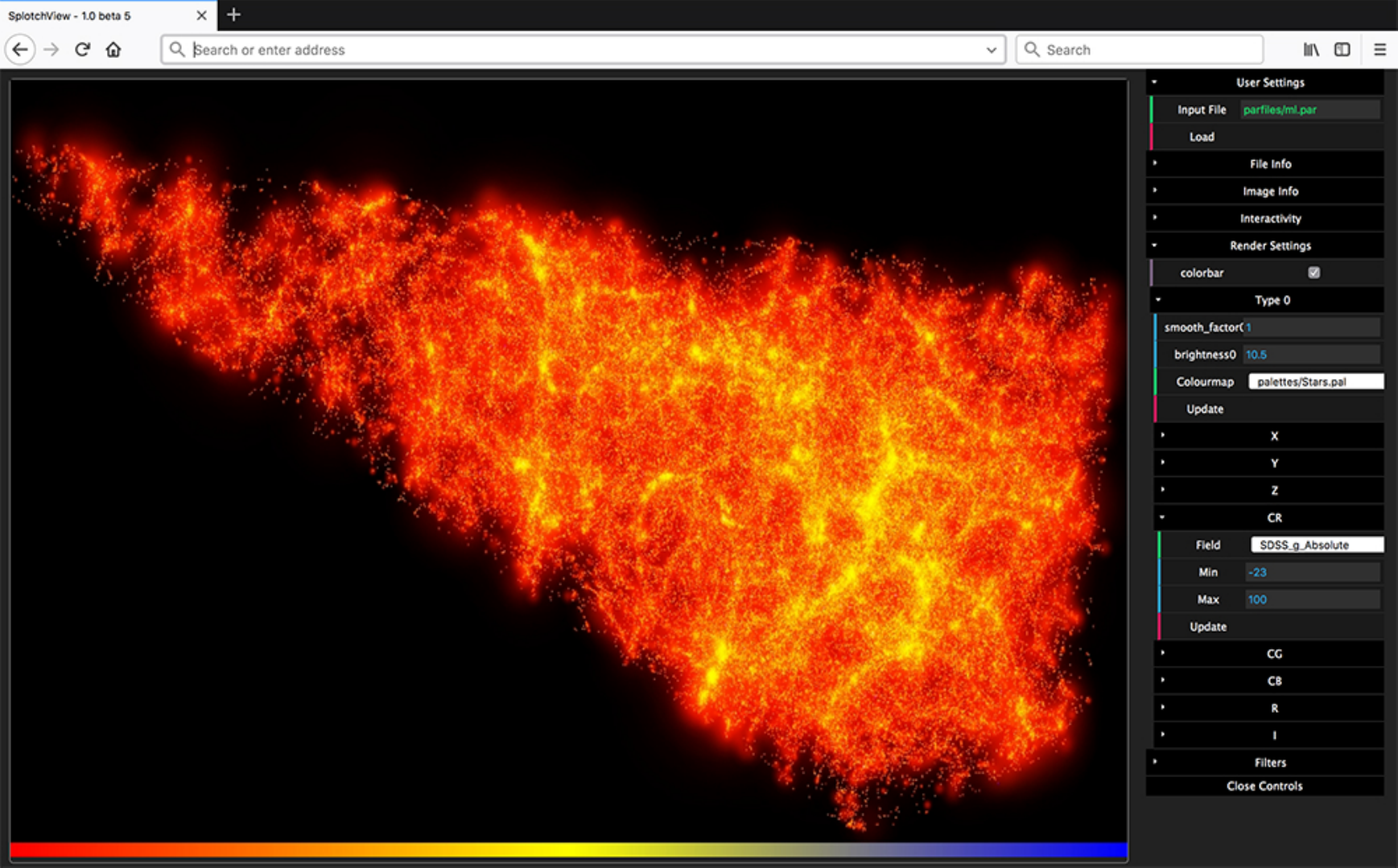}
     \caption{Splotch: A 20 deg\textsuperscript{2} lightcone where \emph{0 < z < 0.1}. Built using the TAO lightcone science module on top of the Millennium with spectral energy distribution computed for SDSS bandpass filters using the model of \citet{conroyetal09}. Galaxies are colouring according to SDSS Absolute G field, while size and intensity are scaled by total stellar mass for visual emphasis.}
     \label{fig:TargetSelection0}
 \end{figure*}

Figure \ref{fig:TargetSelection0} exhibits a 20 deg\textsuperscript{2} lightcone where \emph{0 < z < 0.1}. Built using the lightcone science module on top of the Millennium simulation and SAGE galaxy model, TAO calculates the spectral energy distribution for SDSS bandpass filters using the model of \citet{conroyetal09}. Galaxies are coloured relative to SDSS absolute \emph{g} filter, while size and intensity are further scaled by total stellar mass for visual emphasis. Using a single compute node of our test system Swan, we are able to visualize the full volume of 136298 galaxies interactively at >30 FPS.

\begin{figure*}
 	\includegraphics[width=0.8\textwidth]{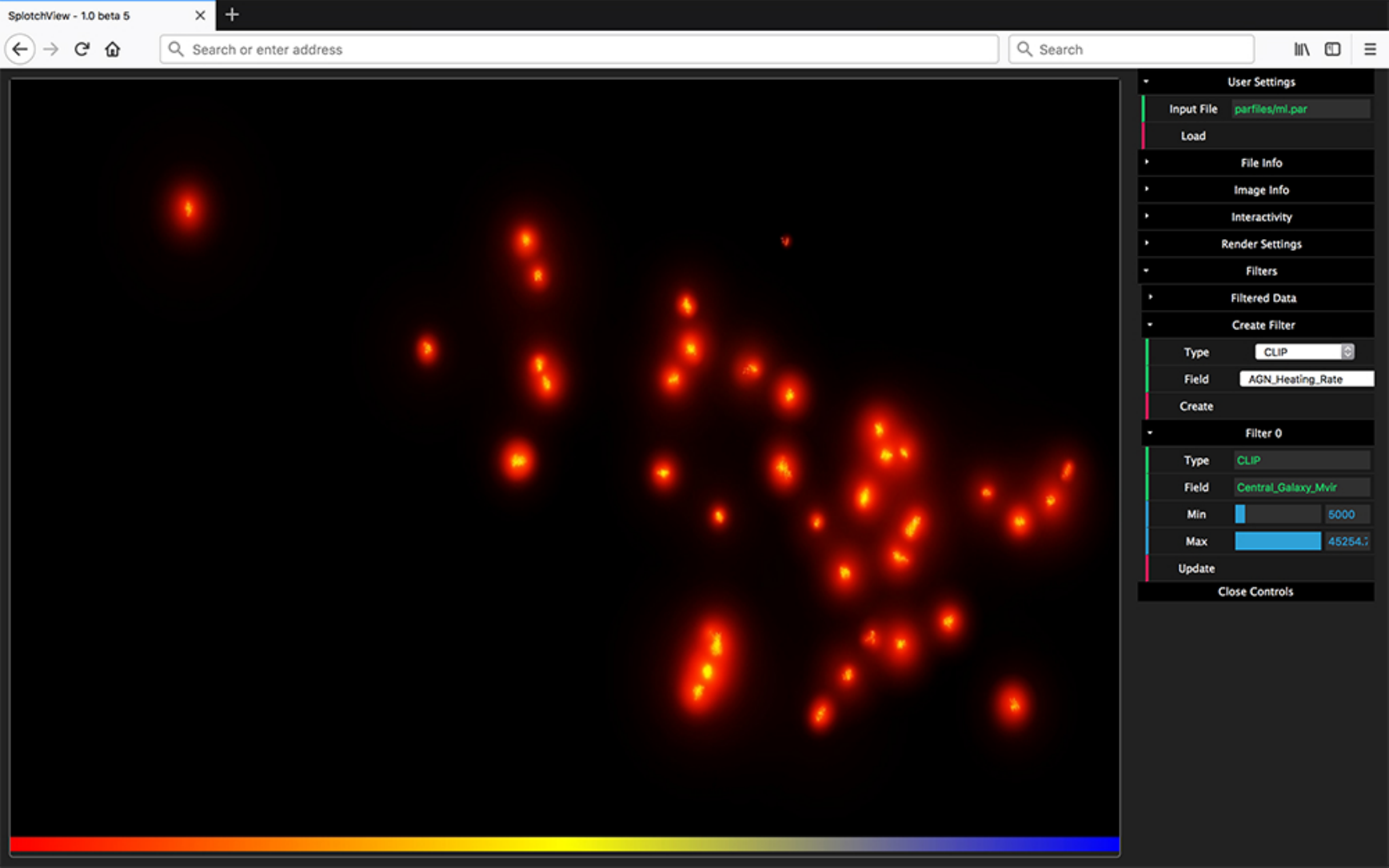}
    \caption{Splotch: the lightcone of Figure \ref{fig:TargetSelection0} is filtered such that only galaxies with central galaxy mVir > 5x10\textsuperscript{13} remain.}
     \label{fig:TargetSelection1}
     
 	\includegraphics[width=0.8\textwidth]{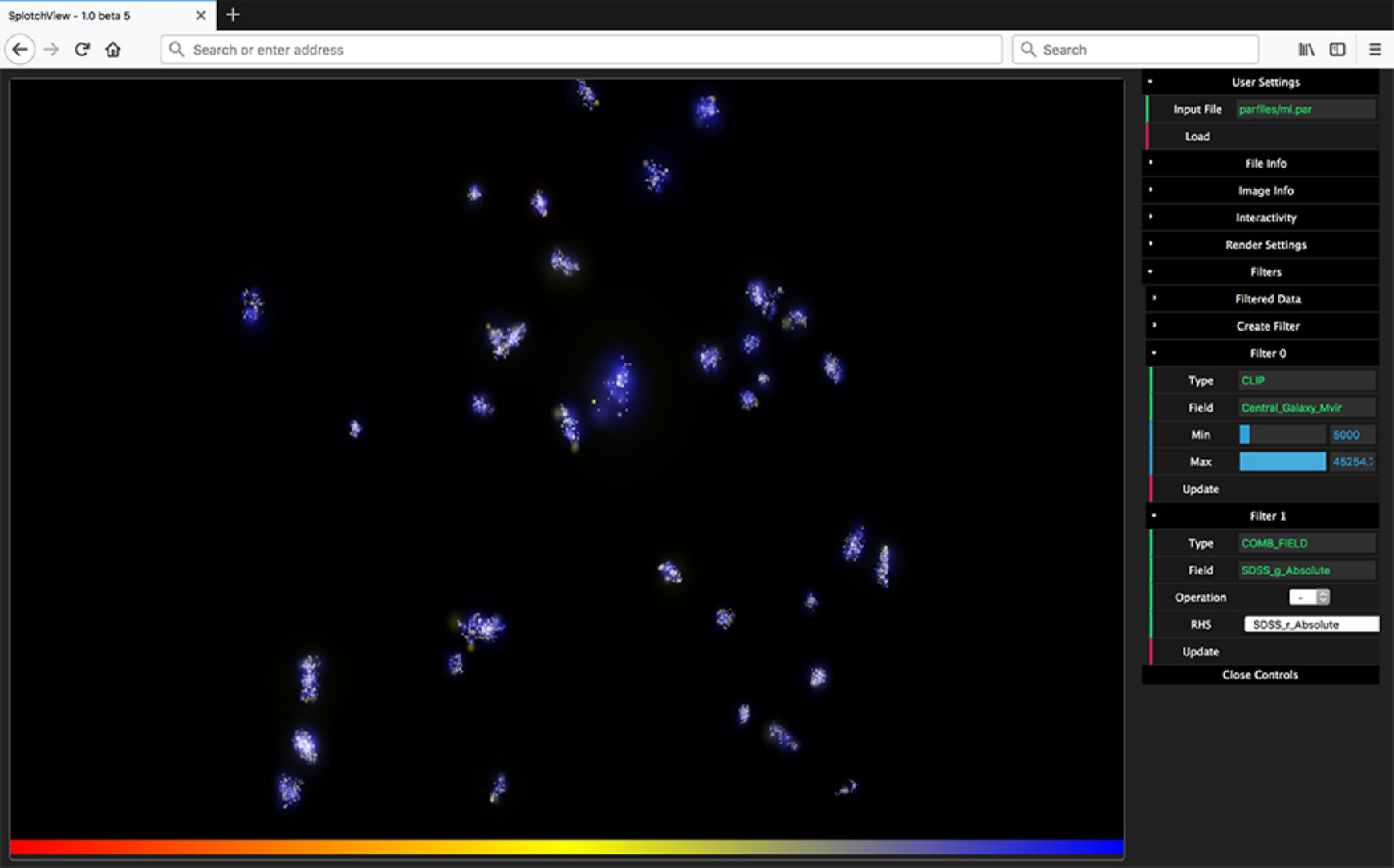}
     \caption{Splotch: a combinatory filter is applied to the remaining galaxies of Figure \ref{fig:TargetSelection1}. The visualized field represents (G-R), while the view has been rotated to align with the direction of increasing redshift. }
     \label{fig:TargetSelection2}
 \end{figure*}

To investigate the rare but significant superclusters in the galaxy distribution, we create a filter that extracts only galaxies where mVir > 5x10\textsuperscript{13} h\textsuperscript{-1}\(\textup{M}_\odot\), shown in Figure \ref{fig:TargetSelection1}. Following this, Figure \ref{fig:TargetSelection2} demonstrates the use of a combinatory filter, colour-coding galaxies by the difference between SDSS bands \emph{g} and \emph{r}. This further highlights the age distribution of galaxies within such massive structures, where bluer galaxies (which tend to be younger and have higher star formation rates) live preferentially on the outskirts, while the redder galaxies (and hence older and mostly devoid of star formation) occupy the cluster cores. The series of figures demonstrate an interactive and in-situ process where the user can find unique populations, such as galaxy clusters and the distribution of galaxies within such structures, which can then be imaged without the need for external analysis. The subsequent output, images and data, can then be jointly downloaded to local storage for subsequent use and publication.

\section{Discussion and Next Steps}
\label{sec:discussion}

Interactive visualization can provide an added value to the process of extracting theoretical datasets from newly emerging web portals. As demonstrated in Section ~\ref{sec:appsincontext}, some of the common scenarios in which a scientist may require data from a theoretical web portal can be accomplished in a more fluid manner with visualization and interactive data filtering techniques. We have demonstrated a client-server tool for interactive web visualization, which we see as an embeddable module within a more general web framework. We are aiming to apply this work in multiple facilities, initially targeting TAO. However, there are further challenges to be addressed before the system is integrated in a production facility, such as the integration of interactive jobs to the scheduling system of TAO. As discussed previously, the traditional model of running jobs on HPC systems is batch. For interactive visualization to work effectively in this model, the user and scheduling system must be able to cooperate to schedule the resources at a time that is acceptable. However, it is becoming more common for high performance systems to have a variety of queues for varying job size, hardware, and memory requirements. It is starting to become common to include interactive queue with dedicated resources, precisely for jobs such as visualization, debugging and other interactive tasks. In this model it is feasible to imagine the ideal scenario where the user requests an interactive session and receives it instantly.

Furthermore, it must be considered that users of the TAO service are likely not also users of the underlying HPC service, and as such they should not be able to exploit the rendering server in order to access restricted data. In consideration of this, Splotch contains a compile-time option to restrict the ability to load files direct from a file path on a per-user basis. Instead, the user may only loading datasets from a specific list which can be provided directly by the web framework.

As discussed in Section \ref{sec:webvizastro}, there are a few existing tools that could enable the type of usage we discuss here. Paraview Web is the most advanced tool for connecting HPC and web visualization, and has already been proven useful within the context of PDACS (see Section \ref{sec:webobservatories}). A key difference in Splotch is the optimized support for high quality rendering of particle data using large and heterogeneous supercomputing systems. We envision Splotch as a tool that can be used entirely in its own right as discussed in this paper, or alternatively embedded within a larger and more generic visualization package. For example, Splotch is already exploited as a batch rendering mode within VisIVO, and there has been some work toward building a Splotch plugin for Paraview. We aim to be considered a lightweight alternative to more general visualization software packages, ideal for building into a web module of an online virtual observatory.

The work we present here shows the viability and efficacy of remote visualization as a support for theoretical data web portals, and the functionalities for data processing can already be useful for scientific applications (Section \ref{sec:appsincontext}). The next steps are full integration in a production service, along with further investigation of advanced visualization methods in this context towards a future vision of a user interactively and in real-time making virtual observations as a part of their web-based analysis workflow. 

In relation to this, a further step to be taken is regarding comparative visualization. Along with quantitative methods, comparative approaches to visualization have been used to good effect for tasks such as verifying scientific simulations \citep{Ahrensetal10} and analysing large sets of multi-dimensional data \citep{Vohletal16}. Whilst 3d visualization is inherently suited to viewing of theoretical astronomical data, many astronomers, particularly those from observational backgrounds, are more accustomed to viewing virtual observations. These provide one of the crucial links between theoretical and observational astronomy, and a common factor amongst each of the portals reviewed in Section ~\ref{sec:webobservatories} is the inclusion of virtual observations. 

An experimental science module currently available in TAO is the mock imaging module, which employs the SkyMaker software package \citep{BertinEtal09} to generate mock telescope images considering common observational phenomena such as aperture, optical defects, and others not commonly used in 3d visualization. These images facilitate understanding of the relationship between properties of galaxies and real observations, and can be used to compare synthetic datasets from different sources with each other and telescope observations.

Mock imaging is traditionally not an interactive process. The user sets initial parameters and provides input data, then launches a job to generate an image. We would like to tie this process to the 3d visualization module, such that a user can interactively filter their data in 3d, then launch an on-the-fly mock image generator to see the effects of their interaction in a more recognizable format. While mock image generation need not be as interactive as 3d visualization, the viability of this approach depends on a fast mock imaging tool that can generate images in a reasonable time frame. As such, future work includes the evaluation of mock imaging tools and their suitability for this use case.

\section{Conclusions}
\label{sec:conclusions}

We present an overview of the applications and potential of 3d visualization in astronomy web portals, initially through an examination of methods currently in use in web portals for astronomy, followed by a further examination of the current status of web-capable 3d visualization tools for astronomy. We show a clear lack of 3d visualization capabilities in current astronomy data portals, and additionally only few 3d visualization tools for astronomy that are web-capable.

We further present an approach based on the high performance batch visualization software Splotch, enabled for interactive 3d visualization via the web, and demonstrate an application of our work through the Theoretical Astrophysical Observatory, a virtual observatory for theory data. We comment on future directions for quantitative and comparative visualization to aid the scientific data access workflow.

\section*{Acknowledgements}

We gratefully acknowledge Swinburne Centre for Astrophysics and Supercomputing for hosting author Tim Dykes while the majority of this work was completed. This work utilized the gSTAR national facility at Swinburne University of Technology. gSTAR is funded by Swinburne and the Australian Government's Education Investment Fund.
Thanks to Cray UK for providing further HPC resources used for development, performance, and debugging.




\bibliographystyle{mnras}
\bibliography{biblio} 

\bsp	
\label{lastpage}
\end{document}